\documentclass[aps,prd,twocolumn,superscriptaddress,nofootinbib]{revtex4} 

\usepackage{amssymb}
\usepackage{amsmath}
\usepackage{amssymb}
\usepackage{graphicx}
\usepackage{subfigure}
\usepackage{color}
\usepackage{mathrsfs}
\usepackage{float}
\usepackage[normalem]{ulem}
\usepackage[dvipsnames]{xcolor}

\usepackage[breaklinks=true,colorlinks=true]{hyperref}
\hypersetup{colorlinks=true,citecolor=blue,linkcolor=blue,urlcolor=blue}

\usepackage[utf8]{inputenc}
\usepackage[english]{babel}

\begin{document}

\title{Deformation of sine-Gordon two-soliton solutions in $\varphi^4$ kink-antikink configurations}

\author{Aliakbar Moradi Marjaneh}
\email{moradimarjaneh@iau.ac.ir; moradimarjaneh@gmail.com}
\affiliation{Alikhanyan National Laboratory (Yerevan Physics Institute), 2 Alikhanyan Brothers Street, Yerevan 0036, Armenia}
\affiliation{Department of Physics, Qu.C., Islamic Azad University, Quchan, Iran.}
\author{Danial Saadatmand}
\email{saadatmand.d@gmail.com}
\affiliation{Alikhanyan National Laboratory (Yerevan Physics Institute), 2 Alikhanyan Brothers Street, Yerevan 0036, Armenia}
\author{Fabiano C. Simas}
\email{fc.simas@ufma.br}
\affiliation{Coordena\c c\~ao do Curso de F\'isica - Bacharelado \& Programa de P\'os-Gradua\c c\~ao em F\'isica, Universidade Federal do Maranh\~ao
(UFMA), Campus Universit\'ario do Bacanga, 65085-580, S\~ao Lu\'is, Maranh\~ao, Brazil.}
\author{Jarah Evslin}
\email{jarah@impcas.ac.cn}
\affiliation{Institute of Modern Physics, NanChangLu 509, Lanzhou 730000, China}
\affiliation{University of the Chinese Academy of Sciences, YuQuanLu 19A, Beijing 100049, China}
\author{Dionisio Bazeia}
\email{bazeia@fisica.ufpb.br, dbazeia@gmail.com}
\affiliation{Department of Physics, Federal University of Paraíba, João Pessoa, Paraíba, Brazil.}

\begin{abstract}

In this work, we construct analytical kink–antikink (\(K\bar K\)) configurations in the non-integrable \(\varphi^4\) model by mapping exact solutions of the integrable sine-Gordon system via a field deformation. This procedure yields two distinct classes of configurations, parametrized by the initial half-separation and velocity. We compare these profiles with the standard additive superposition Ansatz in terms of vacuum structure and equation-of-motion residuals, deriving explicit closed-form expressions for the corresponding integrated squared residuals. For small separations, the mapped soliton–antisoliton configurations exhibit appreciably smaller residuals than the naive superposition Ansatz, which in turn performs better than the mapped two-soliton configurations. Although the mapped fields are not exact solutions of the \(\varphi^4\) equation of motion, they provide mathematically consistent, topologically sound, and physically motivated initial data for numerical studies of kink–antikink scattering and resonance phenomena.

\end{abstract}




\maketitle

\section{Introduction}
\label{sec:intro}
Nonlinear field theories provide fundamental frameworks for describing localized, stable, and particle-like coherent structures across diverse domains, including high-energy physics, condensed matter systems, and nonlinear optics~\cite{manton2004topological,vachaspati2006solitons,Kevrekidis.book.2019,Bishop.PhysD.1980}. In $(1+1)$-dimensional space-time, topological solitons known as kinks interpolate between degenerate, topologically distinct vacuum states, with their stability safeguarded against dissipative decay by conserved topological indices~\cite{bazeia2003kinks,Evslin.JHEP.2024}. Because of their localized energy density and robust topological identity, kink configurations serve as archetypal systems for exploring collision dynamics, non-perturbative energy transfer, and the formation of long-lived metastable excitations~\cite{campbell1983resonance,Makhankov.PhysRep.1978}.

A fundamental paradigm in this context is the integrable sine-Gordon (sG) model, which admits exact analytical multi-soliton solutions, including propagating kinks, antikinks, and bound breathers~\cite{rajaraman1982solitons,Dmitriev.Nonlinearity.2000,Dmitriev.PRE.2008,Moradi.EPJB.2018}. Due to the presence of an infinite hierarchy of conserved charges, kink collisions in the sG system are purely elastic.  In sharp contrast, non-integrable scalar theories, prominently the $\varphi^4$ model and higher-order polynomial models such as $\varphi^6$ and double sine-Gordon~\cite{Lohe.PRD.1979,Campbell.PhysD.1983,Gani.EPJC.2018,Saadatmand.EPJB.2022,Adam.PRD.2022}, exhibit significantly richer and more complex phenomenologies. In the $\varphi^4$ model, each isolated kink supports a discrete localized internal vibrational (shape) mode within its linearized perturbation spectrum~\cite{campbell1983resonance,belova1997soliton}. The resonant exchange of energy between the translational zero-mode and this discrete internal shape mode during collisions gives rise to fractal resonance structures, multi-bounce escape windows, and the formation of localized, long-lived oscillating configurations (oscillons)~\cite{campbell1983resonance,anninos1991fractal,sugiyama1979kink,Askari.CSF.2020,Manton.PRL.2021,Almeida.EPJC.2025,Saadatmand.CSF.2024}, {which can be viewed as a generalization of the breathers found in integrable systems}. 

Despite decades of intense investigation, exploring multi-kink dynamics in non-integrable theories remains fundamentally challenged by the absence of exact closed-form multi-soliton solutions. {Pioneering semiclassical approaches by Dashen, Hasslacher, and Neveu (DHN)~\cite{dashen1975semiclassical} explored analytical approximations via mapped breather-like solutions, while at large separations ($d \gtrsim 2$), the asymptotic inter-kink dynamics are effectively governed by asymptotic force expansions as formulated by Manton and others~\cite{manton2004topological, Manton.PRL.2021}, where overlap effects decay exponentially. In sharp contrast, in the truly nonlinear, strong-overlap regime ($d \lesssim 1.5$) at sub-relativistic and moderate velocities, such asymptotic frameworks break down.} Historically, numerical and analytical studies of kink--antikink ($K\bar{K}$) interactions have predominantly relied on the standard linear superposition Ansatz, wherein single-kink profiles are superposed with an initial velocity and separation~\cite{Moradi.CNSNS.2017,Manton.PRL.2021,Christov.PRL.2019}. While kinematically straightforward, this linear heuristic violates the intrinsically nonlinear field equations, generating spurious initial field gradients and unphysical radiation, particularly at small to intermediate separations where kink tails overlap significantly~\cite{Campos.JHEP.2024,Andre.AnlPhys.2025,Campos.JHEP.2025}. Alternative formulations, such as variational collective coordinate models~\cite{sugiyama1979kink,campbell1983resonance,Saadatmand.2014.dynamics,Weigel.PRD.2016,Pereira.JoPA.2021}, capture key dynamical degrees of freedom but can suffer from coordinate singularities or truncation artifacts.

A powerful pathway to bridge the gap between integrable and non-integrable theories is provided by the \textit{deformation procedure}~\cite{bazeia2002deformed,bazeia2006deformation}. By employing a differentiable deformation map $f(\varphi)$, one can systematically map exact multi-soliton solutions of an integrable system (such as the sine-Gordon theory) onto the field space of a target non-integrable model (such as the $\varphi^4$ model). These mapped profiles provide physically well-motivated and structurally coherent initial data for exploring the subsequent nonlinear collision dynamics, resulting from mapping multi-soliton configurations of an integrable model (such as the sG theory) onto the field space of a target non-integrable theory (such as the $\varphi^4$ model). Unlike heuristic superposition prescriptions, deformed configurations preserve the structural nonlinearity of the underlying field transformations by construction. Although these mapped configurations do not satisfy the dynamic equations of motion of the target non-integrable model identically, they provide highly consistent and physically motivated field profiles that mitigate unphysical overlap distortions.

In this work, we present a systematic analytical construction and quantitative benchmarking of kink--antikink ($K\bar{K}$) configurations in the $\varphi^4$ model, obtained via the sG-to-$\varphi^4$ deformation map. To evaluate the mathematical and dynamical fidelity of these constructions, we perform a comprehensive residual analysis that quantitatively benchmarks three distinct configurations: the conventional superposition Ansatz and two distinct classes (Class~I and Class~II) of deformed solutions. We establish explicit parametric matching conditions that link the deformation parameters directly to the physical half-separation $d$ and velocity $v$ at the {approach-phase} reference time $t_0(d) = -d/v$. By analyzing both local field equation residuals and explicit closed-form integrated squared residuals $\mathcal{R}(d,v)$, we delineate the regimes where deformed configurations offer substantial fidelity advantages over the standard superposition Ansatz, providing an efficient, closed-form analytical description for strongly interacting kink-antikink configurations at close separations ($d \lesssim 1.5$) where standard superposition fails, with potential applications to dense soliton systems and moduli-space approximations.

The remainder of this paper is structured as follows. In Sec.~\ref{sec:deformation}, we review the mathematical framework of the deformation method connecting the sine-Gordon and $\varphi^4$ scalar theories. In Sec.~\ref{sec:configurations}, we construct the explicit analytical forms of the deformed $K\bar{K}$ configurations, formulate the parametric matching conditions, and derive the analytical expressions for the equation-of-motion residuals. In Sec.~\ref{sec:results}, we present the comparative benchmarking of the local and integrated squared residuals across varying separations and velocities. Finally, Sec.~\ref{sec:conclusion} summarizes our conclusions and outlines future applications to collision dynamics. The closed-form analytical expressions and auxiliary polynomials for the integrated residuals are detailed in Appendix~\ref{app:integrated_squared_residual}.

\section{%
\texorpdfstring{Deformation Method and the $\varphi^4$ Model}
  {Deformation Method and the phi4 Model}%
}\label{sec:deformation}

The deformation procedure provides a systematic framework for constructing scalar field theories and associated field configurations from a known starting model \cite{bazeia2002deformed,bazeia2006deformation}. Its central idea is to
map exact solitons of an integrable theory into configurations of a target theory through a suitable differentiable transformation. This approach preserves analytic control over the resulting field profiles, while the accuracy of the mapped configurations can be assessed using the equation of motion of the target theory.

Let $\chi(x,t)$ denote the field of the integrable model, governed by the
potential $V(\chi)$. Given a differentiable deformation function
$f(\varphi)$ with a nonvanishing derivative, the potential of the deformed
theory, $\widetilde{V}(\varphi)$, is defined as
\begin{equation}
\widetilde{V}(\varphi) = \frac{V\!\left(f(\varphi)\right)}{\left[f'(\varphi)\right]^2}.
\label{eq:deformed_potential}
\end{equation}
If $\chi_{\mathrm{sol}}(x,t)$ is an exact solution of the original equation of
motion, the corresponding configuration in the deformed model is obtained
through the inverse map
\begin{equation}
\varphi(x,t) = f^{-1}\!\left[\chi_{\mathrm{sol}}(x,t)\right].
\label{eq:deformed_kink}
\end{equation}

In this work we take the sine-Gordon (sG) model as the starting point, with
potential
\begin{equation}
V^{\mathrm{sG}}(\chi) = 1 - \cos\chi,
\label{eq:sg_potential}
\end{equation}
whose equation of motion, $\chi_{tt} - \chi_{xx} + \sin\chi = 0$, is integrable.
The model admits the well-known moving kink ($+$) and antikink ($-$) solutions
\begin{equation}
\chi^{\mathrm{sG}}_{K,\bar K}(x,t) = 4\tan^{-1}\!\left[ e^{\pm \gamma (x-vt)} \right],
\label{eq:sg_kink}
\end{equation}
with $\gamma = (1-v^2)^{-1/2}$ the Lorentz factor.

To map these solutions onto the $\varphi^4$ theory we use the specific
deformation function
\begin{equation}
f(\varphi) = 4\tan^{-1}\!\left( e^{c/2} \sqrt{\frac{1+\varphi}{1-\varphi}} \right),
\label{eq:deformation_function}
\end{equation}
where $c$ is a real deformation parameter that controls the relative
positioning and shape of the resulting kink-antikink configuration. Its inverse,
which maps $\chi \in (0,2\pi)$ onto $\varphi \in (-1,1)$ and extends
continuously to the vacuum limits, reads
\begin{equation}
f^{-1}(\chi) = \frac{\tan^2(\chi/4) - e^c}{\tan^2(\chi/4) + e^c}.
\label{eq:inverse_deformation}
\end{equation}

Substituting Eqs.~\eqref{eq:sg_potential} and~\eqref{eq:deformation_function}
into Eq.~\eqref{eq:deformed_potential}, the deformed potential reduces exactly
to the standard $\varphi^4$ double-well potential, vanishing at its two vacua:
\begin{equation}
V(\varphi) = \frac{1}{2}\left(\varphi^2-1\right)^2.
\label{eq:phi4_potential}
\end{equation}
The associated equation of motion is
\begin{equation}
\varphi_{tt} - \varphi_{xx} + 2\varphi(\varphi^2-1) = 0.
\label{eq:phi4_eom}
\end{equation}

As a consistency check, consider the static sine-Gordon kink
$\chi(x)=4\tan^{-1}(e^x)$. For $c=0$, substitution into
Eq.~\eqref{eq:inverse_deformation} gives the exact $\varphi^4$ kink solution,
\begin{equation}
\varphi(x)=\tanh x.
\end{equation}
The corresponding boosted kink and antikink profiles are
\begin{equation}
\varphi_{K,\bar K}(x,t) = \pm \tanh\!\left[\gamma(x-vt)\right].
\label{eq:phi4_kink_moving}
\end{equation}

It is important to stress that, although the sG model is integrable and 
possesses exact multi-soliton solutions, the $\varphi^4$ theory is 
non-integrable. Therefore, when the deformation method is applied to sG two-soliton configurations, 
the resulting $\varphi^4$ profiles should be interpreted as analytically constructed 
approximations of kink-antikink configurations. While these profiles 
do not satisfy the $\varphi^4$ equations of motion identically, they 
provide a physically motivated starting point for studying nonlinear 
dynamics, rather than exact solutions of the full $\varphi^4$ evolution problem. 
Their deviation from Eq.~\eqref{eq:phi4_eom} is quantified by the residual analysis 
carried out in the following sections, where these {deformed 
approximations} are compared with the {standard linear 
superposition method} at the level of spacetime profiles, local residuals, 
and integrated squared residuals. In Sec.~\ref{subsec:comparison_matching}, the 
parameter $c$ will be fixed separately for each {approach} 
as a function of the kink separation and velocity through the parametric 
matching condition introduced below.

\section{Configurations}
\label{sec:configurations}

The periodic vacuum structure of the sine-Gordon model allows for an infinite number of topological sectors. However, under the deformation map to the $\varphi^4$ theory, these solutions are projected onto a model with only two degenerate vacua, $\varphi=\pm 1$. Consequently, the deformation procedure generates two distinct classes of kink--antikink ($K\bar{K}$) configurations in the $\varphi^4$ theory, depending on the initial sine-Gordon sector.

\subsection{Deformed Configurations: Class I and Class II}
\label{subsec:deformed_configs}

The first class, denoted as Class I, originates from the sine-Gordon $K\bar{K}$ sector. The exact sG solution is given by \cite{rajaraman1982solitons}:
\begin{equation}
\chi^{\mathrm{sG}}_{K\bar{K}}(x, t) = 4 \arctan\left( \frac{\sinh(\gamma v t)}{v \cosh(\gamma x)} \right).
\label{eq:sg_kak}
\end{equation}

Applying the inverse deformation map \eqref{eq:inverse_deformation} and simplifying via hyperbolic identities, we obtain the $\varphi^{(I)}$ configuration, {which represents the full time-dependent profile}:
\begin{equation}
\varphi^{(I)}(x, t) = \frac{ \sinh^2(\gamma v t) - e^c v^2 \cosh^2(\gamma x) }{ \sinh^2(\gamma v t) + e^c v^2 \cosh^2(\gamma x) }.
\label{eq:phi4_deformed_I}
\end{equation}

The second class, Class II, is generated from the sine-Gordon kink--kink ($KK$) sector \cite{rajaraman1982solitons}:
\begin{equation}
\chi^{\mathrm{sG}}_{KK}(x, t) = 4 \arctan\left( \frac{v\sinh(\gamma x)}{\cosh(\gamma v t)} \right).
\label{eq:sG_kk}
\end{equation}
Deforming this solution yields the $\varphi^{(II)}$ configuration:
\begin{equation}
\varphi^{(II)}(x, t) = \frac{ e^c \cosh^2(\gamma v t) - v^2 \sinh ^2(\gamma x) }{ e^c \cosh^2(\gamma v t) + v^2 \sinh ^2(\gamma x) }.
\label{eq:phi4_deformed_II}
\end{equation}

It is noteworthy that $\varphi^{(I)}$ and $\varphi^{(II)}$ originate from
different sine-Gordon sectors: Class~I is obtained from the topologically
trivial $K\bar{K}$ sector, whereas Class~II is generated from the
$Q=2$ topological $KK$ sector. Nevertheless, the deformation map projects
both configurations onto the topologically trivial sector of the
$\varphi^4$ theory. In both cases, the field satisfies
\[
\varphi(\pm\infty,t)=-1,
\]
while the interior region approaches the second vacuum,
$\varphi=+1$, as the two-kink structure becomes more separated. Although
$c$ reduces to a spatial translation for an isolated kink, it enters
nontrivially into the two-kink profiles through the inverse deformation map
and is fixed here by the parametric matching conditions discussed in
Sec.~\ref{subsec:comparison_matching}.

\subsection{Kink--Antikink Separation and the Standard Superposition Prescription}
\label{subsec:superposition}

{To demonstrate the advantages of the deformation-based approach, we compare it against the widely used standard additive superposition prescription. This conventional method, frequently employed in scattering studies, serves as a baseline for evaluating the fidelity of the analytically constructed configurations.} 
For the {approach phase}, $t<0$, the configuration is defined by
\begin{equation}
\varphi_{\mathrm{sup}}(x,t)=
-\tanh[\gamma(x+vt)]
+\tanh[\gamma(x-vt)]
-1,
\label{eq:sup-full-in}
\end{equation}
while for the {separation phase}, $t\ge 0$, it is given by
\begin{equation}
\varphi_{\mathrm{sup}}(x,t)=
-\tanh[\gamma(x-vt)]
+\tanh[\gamma(x+vt)]
-1.
\label{eq:sup-full-out}
\end{equation}

The centers of the individual kinks are located at $x_{1,2} = \pm v|t|$. We define the half-separation as $d = v|t|$, such that the total distance between centers is $2d$. To compare different configurations at a prescribed separation $2d$ and velocity $v$, we define a reference time:
\begin{equation}
t_0(d) = -d/v.
\label{eq:t0-d}
\end{equation}
Evaluating \eqref{eq:sup-full-in} at $t_0(d)$ yields the static snapshot of the superposition:
\begin{equation}
\varphi_{\mathrm{sup}}(x;d,v) = -\tanh\!\left[\gamma(x-d)\right] + \tanh\!\left[\gamma(x+d)\right] - 1.
\label{eq:superposition-d}
\end{equation}
This formulation ensures that all three configurations
($\varphi_{\mathrm{sup}}$, $\varphi^{(I)}$, and $\varphi^{(II)}$) can be
compared on an equal footing by identifying $d$ as the common half-separation
parameter. The limit $d\to0^+$ corresponds to the zero-separation limit, in which the
two-kink structure collapses toward the vacuum configuration
$\varphi=-1$ at the reference time.

\subsection{Comparison and Parametric Matching}
\label{subsec:comparison_matching}

The deformation parameters $c_I$ and $c_{II}$ are determined by parametric matching conditions, which express them in terms of the common half-separation $d$ and velocity $v$:
\begin{equation}
e^{c_I(d)}
=
\frac{1}{v^2}\tanh^2(\gamma d),
\qquad
e^{c_{II}(d)}
=
v^2\tanh^2(\gamma d).
\label{eq:matching_conditions}
\end{equation}
The residuals $\mathcal{R}_A(d,v)$ presented in the following sections are evaluated at the reference time $t=t_0(d)$ after this matching procedure.

To evaluate and compare the quality of the three {analytically constructed configurations}, we utilize the local residual $R_A(x;d,v)$ derived from the standard $\varphi^4$ equation of motion:\begin{equation}
R_A(x;d,v)
=
\left[
\partial_t^2\varphi_A
-\partial_x^2\varphi_A
+2\varphi_A\left(\varphi_A^2-1\right)
\right]_{t=t_0(d)},
\label{eq:residual_definition}
\end{equation}
where $\gamma=(1-v^2)^{-1/2}$ and $t_0(d)=-d/v$ is the reference time on the {approach phase}.

For the superposition Ansatz, the local residual is
\begin{equation}
R_{\mathrm{sup}}(x;d,v)
=
\frac{
12\left(1-e^{-4\gamma d}\right)
}{
\left[
\cosh(2\gamma d)+\cosh(2\gamma x)
\right]^2
}.
\label{eq:Rsup}
\end{equation}

For Class~I, the residual can be expressed compactly as
\begin{align}
R_I(x;d,v)
={}&
\frac{4\,\coth^2(\gamma d)}
{
(v^2-1)
\left[
2+\cosh(2\gamma d)+\cosh(2\gamma x)
\right]^3
}
\notag\\
&\times
\Big[
\mathcal{C}_0^{(I)}
+\mathcal{C}_1^{(I)}\cosh(2\gamma x)
-2v^2\cosh^2(2\gamma x)
\Big],
\label{eq:RI}
\end{align}
where
\begin{align}
\mathcal{C}_0^{(I)}
={}&
5+4v^2+\cosh(4\gamma d)
+6(v^2-1)\cosh(2\gamma d),
\label{eq:C0_I}\\
\mathcal{C}_1^{(I)}
={}&
2(3+v^2)
+6(v^2-1)\cosh(2\gamma d).
\label{eq:C1_I}
\end{align}

Similarly, for Class~II, the residual is written as
\begin{align}
R_{II}(x;d,v)
={}&
\frac{
\operatorname{sech}^{6}(\gamma d)
\tanh^{2}(\gamma d)
}{
2(v^2-1)
\left[
\operatorname{sech}^{2}(\gamma d)\sinh^{2}(\gamma x)
+\tanh^{2}(\gamma d)
\right]^3
}
\notag\\
&\times
\Big[
\mathcal{C}_0^{(II)}
+\mathcal{C}_1^{(II)}\cosh(2\gamma x)
+2v^2\cosh^{2}(2\gamma x)
\Big],
\label{eq:RII}
\end{align}
where the corresponding coefficients are
\begin{align}
\mathcal{C}_0^{(II)}
={}&
-\left[
5+4v^2+\cosh(4\gamma d)
\right]
+6(v^2-1)\cosh(2\gamma d),
\label{eq:C0_II}\\
\mathcal{C}_1^{(II)}
={}&
2(3+v^2)
-6(v^2-1)\cosh(2\gamma d).
\label{eq:C1_II}
\end{align}

To quantify the global discrepancy over the entire spatial domain,
we use the integrated squared-residual measure
\begin{equation}
\mathcal{R}_A(d,v)
\equiv
\int_{-\infty}^{+\infty}
\left[R_A(x;d,v)\right]^2\,dx,
\qquad
A\in\{\mathrm{sup},I,II\}.
\label{eq:integrated_residual}
\end{equation}

The matching conditions determine the deformation parameters as
functions of $d$ and $v$. The local residuals
$R_A(x;d,v)$ and the integrated measures
$\mathcal{R}_A(d,v)$ then provide quantitative criteria for comparing
the three configurations as functions of the half-separation $d$ and
the velocity $v$.

For numerical evaluation, the infinite spatial domain is mapped onto
the compact interval $[-1,1]$ through
\[
z=\tanh(\gamma x),
\]
with further details given in Appendix~\ref{app:integrated_squared_residual}.

\section{Numerical Evaluation and Residual Analysis}
\label{sec:results}

In this section, we numerically evaluate the analytical Ansatzes
$\varphi_{\mathrm{sup}}$, $\varphi^{(I)}$, and $\varphi^{(II)}$
and their residuals with respect to the $\varphi^4$ equation of motion.
No numerical time integration of the field equation is performed. The
displayed spacetime surfaces are direct evaluations of the prescribed
analytical profiles, while the residual measures are obtained from the
analytical residual formulas and their spatial integrals.

\subsection{Field Configurations and Spacetime Profiles}

\begin{figure*}[ht!]
\centering
\subfigure[]{\includegraphics[width=0.32\textwidth]{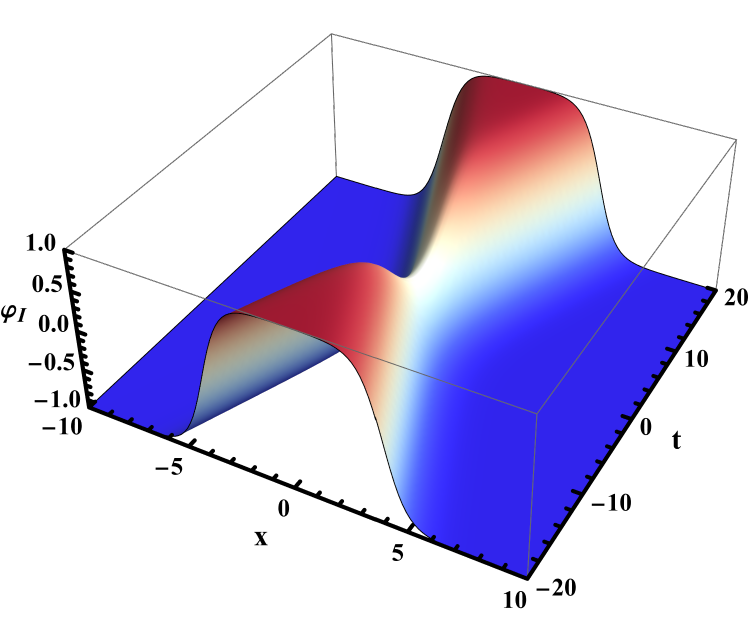}\label{fig:PhiI3Dv0.2}}
\subfigure[]{\includegraphics[width=0.32\textwidth]{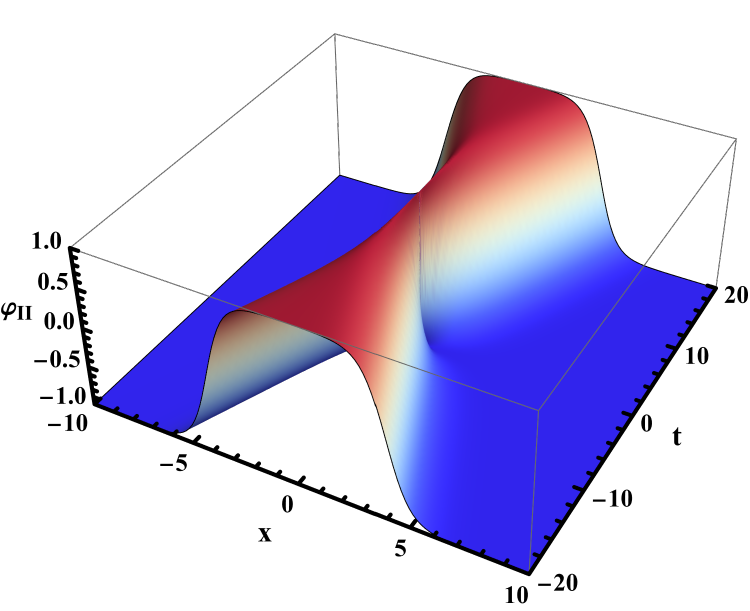}\label{fig:PhiII3Dv0.2}}
\subfigure[]{\includegraphics[width=0.32\textwidth]{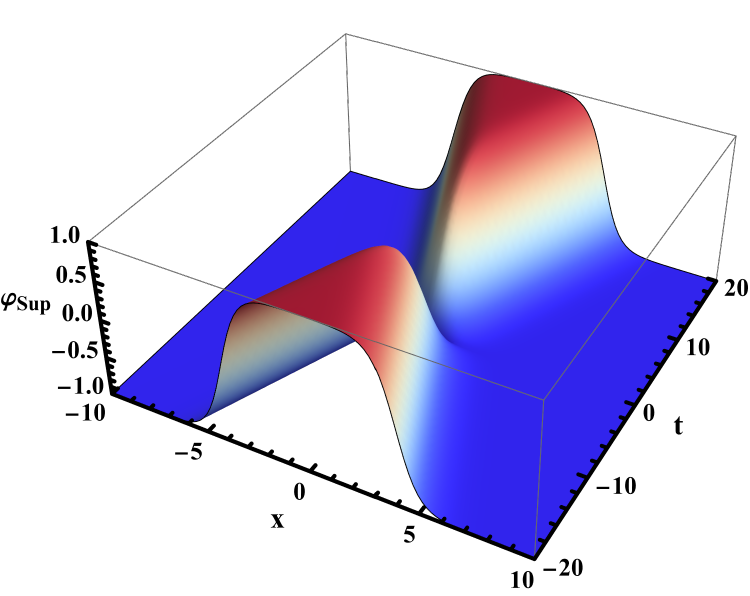}\label{fig:PhiSup3Dv0.2}}
\caption{Three-dimensional representations of the analytical configurations
as functions of $x$ and $t$ at $v=0.2$: (a) the matched Class I profile,
obtained from Eq.~\eqref{eq:phi4_deformed_I}; (b) the matched Class II
profile, obtained from Eq.~\eqref{eq:phi4_deformed_II}; and (c) the
piecewise superposition Ansatz defined by
Eqs.~\eqref{eq:sup-full-in} and \eqref{eq:sup-full-out}. For the deformed
profiles, the instantaneous half-separation is identified as $d=v|t|$, and
the matching conditions in Eq.~\eqref{eq:matching_conditions} are substituted
into Eqs.~\eqref{eq:phi4_deformed_I} and
\eqref{eq:phi4_deformed_II}. The surfaces are direct evaluations of the
prescribed analytical expressions and do not represent solutions obtained
through numerical time integration of the $\varphi^4$ field equation.}
\label{fig:Phi3D}
\end{figure*}

For the spacetime representations, the instantaneous half-separation is
identified as $d=v|t|$. Substituting this relation into the matching
conditions in Eq.~\eqref{eq:matching_conditions} reduces the deformed
profiles to
\begin{align}
\varphi^{(I)}_{\mathrm{m}}(x,t)
&=
\frac{\cosh^2(\gamma vt)-\cosh^2(\gamma x)}
{\cosh^2(\gamma vt)+\cosh^2(\gamma x)},
\label{eq:phi4_matched_I}\\
\varphi^{(II)}_{\mathrm{m}}(x,t)
&=
\frac{\sinh^2(\gamma vt)-\sinh^2(\gamma x)}
{\sinh^2(\gamma vt)+\sinh^2(\gamma x)}.
\label{eq:phi4_matched_II}
\end{align}
Figure~\ref{fig:Phi3D} displays these matched spacetime profiles together
with the piecewise superposition Ansatz at $v=0.2$. These surfaces are direct
evaluations of the prescribed analytical expressions and do not represent
solutions obtained by numerical time integration of the nonlinear field
equation. The matched spacetime profiles in
Eqs.~\eqref{eq:phi4_matched_I} and \eqref{eq:phi4_matched_II} are used
only for the three-dimensional visualization. The field and residual
comparisons below are performed at the reference snapshot
$t=t_0(d)$ for independently prescribed values of $d$.

\begin{figure*}[!ht]
\centering
\subfigure[]{\includegraphics[width=0.32\textwidth]{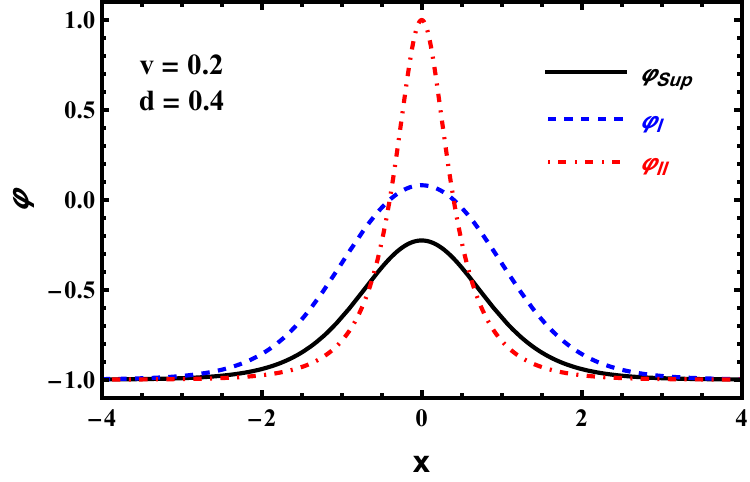}\label{fig:PhiI2Dv0.2d0.4}}
\subfigure[]{\includegraphics[width=0.32\textwidth]{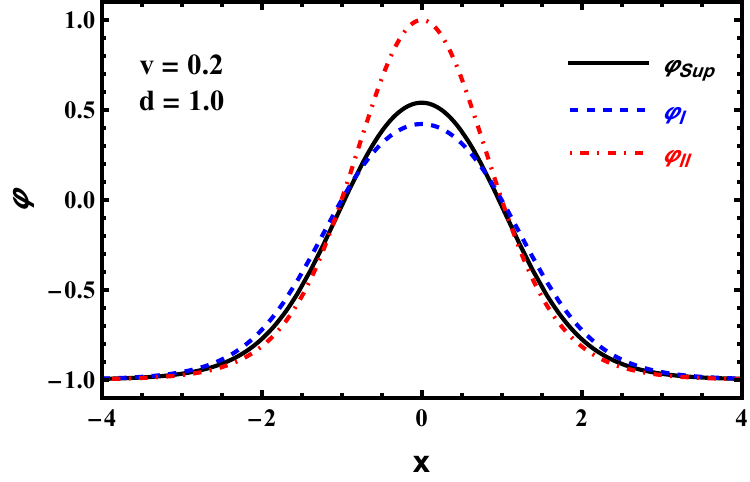}\label{fig:PhiI2Dv0.2d1.0}}
\subfigure[]{\includegraphics[width=0.32\textwidth]{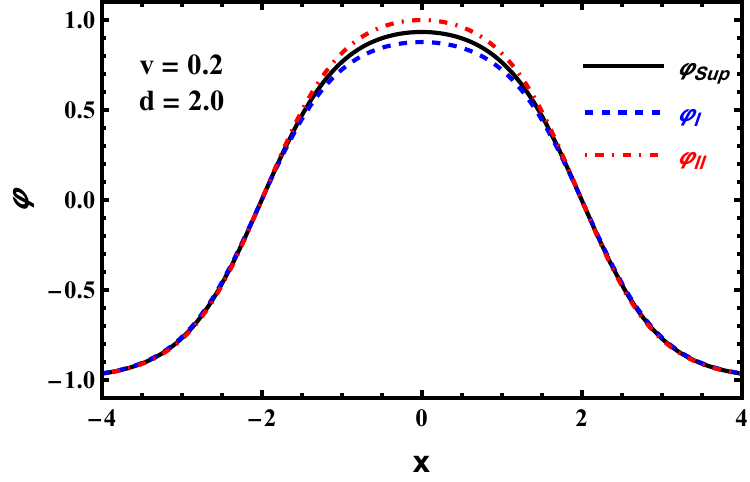}\label{fig:PhiI2Dv0.2d2.0}}
\caption{Field profiles $\varphi(x,t_0)$ evaluated at the \textcolor{RoyalBlue}{approach-phase} time $t_0(d)=-d/v$ for three representative half-separations:
(a) $d=0.4$, (b) $d=1.0$, and (c) $d=2.0$, with fixed velocity
$v=0.2$. In each panel, the black solid, blue dashed, and red dot-dashed
curves denote $\varphi_{\mathrm{sup}}$ in
Eq.~\eqref{eq:sup-full-in}, $\varphi^{(I)}$ in
Eq.~\eqref{eq:phi4_deformed_I}, and $\varphi^{(II)}$ in
Eq.~\eqref{eq:phi4_deformed_II}, respectively. The profiles cover the
strong-overlap, intermediate, and weak-overlap regimes and are displayed
with the same ordering used in the residual analysis.}
\label{fig:Phi2D}
\end{figure*}

To compare the three Ansatzes at the field level, we evaluate them at the {approach-phase} reference time $t_0(d)=-d/v$ [Eq.~\eqref{eq:t0-d}] for
the same prescribed half-separation $d$. Figure~\ref{fig:Phi2D} shows the
resulting profiles $\varphi(x,t_0)$ for three representative values of $d$,
covering the strong-, intermediate-, and weak-overlap regimes. All three
configurations approach the same vacuum, $\varphi\to-1$, as
$x\to\pm\infty$, but they differ noticeably in the central overlap region.
The discrepancy is substantial in the strong- and intermediate-overlap
regimes and becomes increasingly localized near the center as the
separation increases.

\subsection{Local Residual Analysis}

\begin{figure*}[ht!]
\centering
\subfigure[]{\includegraphics[width=0.32\textwidth]{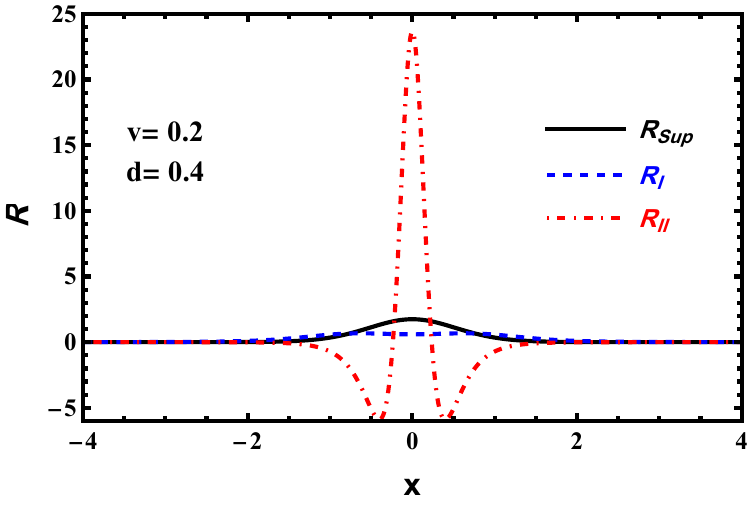}\label{fig:Rv0.2d0.4}}
\subfigure[]{\includegraphics[width=0.32\textwidth]{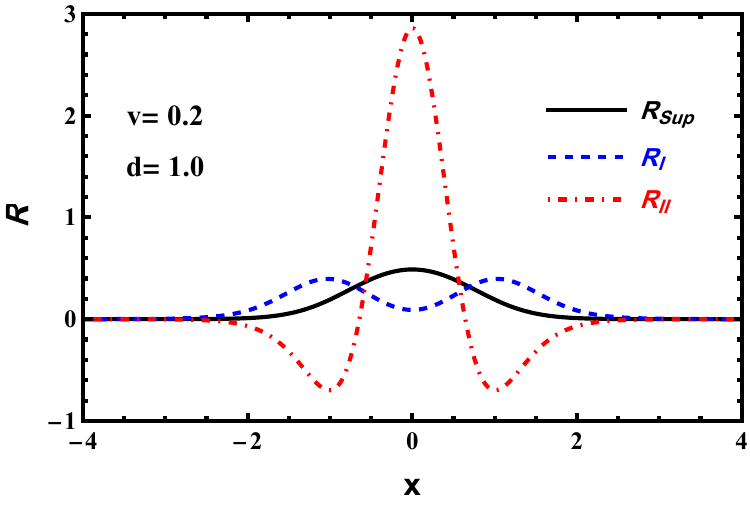}\label{fig:Rv0.2d1.0}}
\subfigure[]{\includegraphics[width=0.32\textwidth]{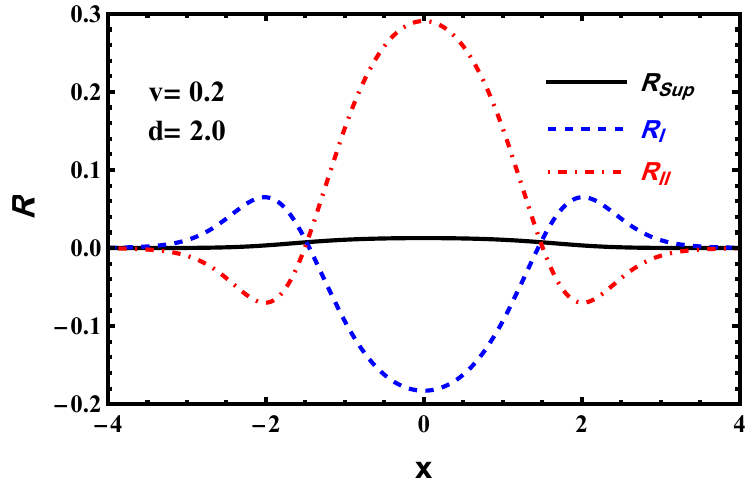}\label{fig:Rv0.2d2.0}}
\caption{Local residual profiles $R_A(x;d,v)$ as functions of $x$ at
$v=0.2$ for three representative half-separations: (a) $d=0.4$,
(b) $d=1.0$, and (c) $d=2.0$. In each panel,
$R_{\mathrm{sup}}$ is represented by the black solid curve,
$R_I$ by the blue dashed curve, and $R_{II}$ by the red dot-dashed
curve. These residuals are defined in Eqs.~\eqref{eq:Rsup},
\eqref{eq:RI}, and \eqref{eq:RII}, respectively. The panels show the
overall reduction of the residual amplitudes as the overlap decreases
and illustrate the distinct spatial structures generated by the three
Ansatzes.}
\label{fig:Residual_Profiles}
\end{figure*}

The accuracy of each configuration is quantified by its local residual
$R_A(x;d,v)$ with respect to the $\varphi^4$ equation of motion,
Eq.~\eqref{eq:residual_definition}. Figure~\ref{fig:Residual_Profiles}
shows the spatial distribution of $R_A$ at $v=0.2$ for the three
representative half-separations. For all configurations, the residual
is spatially localized within the interaction region and rapidly decays to
zero as $|x|\to\infty$, confirming that deviations from the exact field equation
arise predominantly from the nonlinear overlap of the kink tails.

The detailed spatial structure of the residual depends significantly on the
chosen Ansatz and separation. At $d=0.4$, the Class~II residual exhibits a sharp
positive central peak accompanied by shallow negative side lobes, while
$R_{\mathrm{sup}}$ develops a moderate central maximum. In contrast, the Class~I
residual remains remarkably small across the core region on this scale.
As $d$ increases, the residuals exhibit multi-lobed profiles; for instance, at
$d=1.0$, $R_I$ develops off-center extrema rather than a simple central peak.
Overall, the maximum residual amplitude of each Ansatz diminishes significantly
with increasing separation, with the superposition residual becoming smallest
over most of the domain at $d=2.0$.

\subsection{Integrated Squared Residuals and Global Accuracy}

\begin{figure*}[ht!]
\centering
\subfigure[]{\includegraphics[width=0.32\textwidth]{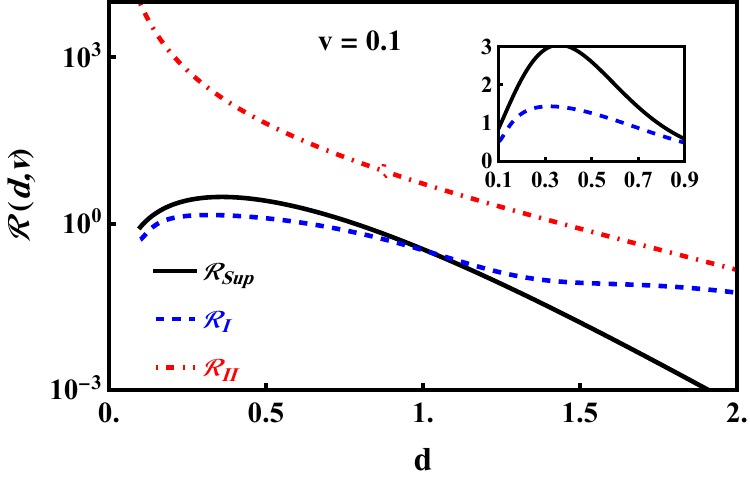}\label{fig:IntRv0.1}}
\subfigure[]{\includegraphics[width=0.32\textwidth]{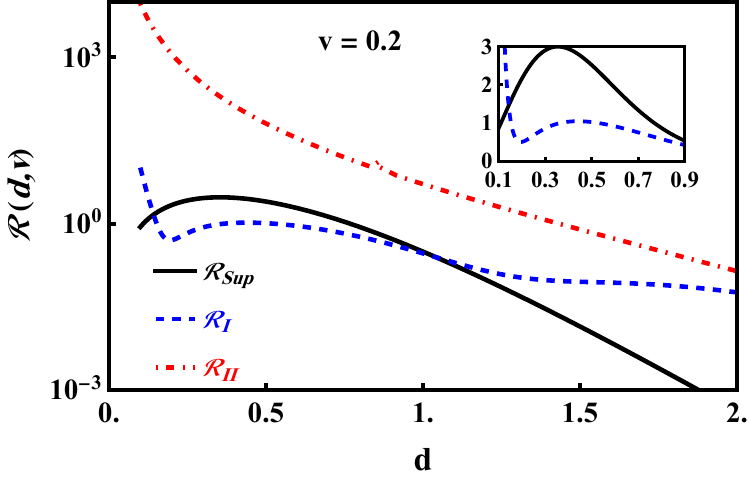}\label{fig:IntRv0.2}}
\subfigure[]{\includegraphics[width=0.32\textwidth]{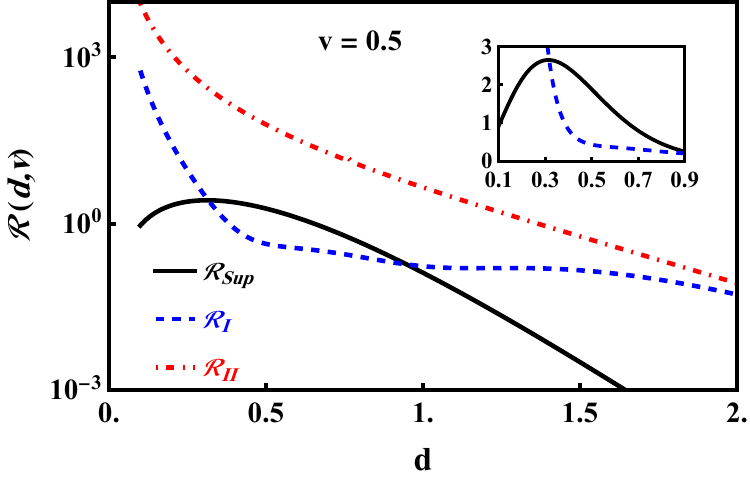}\label{fig:IntRv0.5}}
\caption{Integrated squared residuals $\mathcal{R}_A(d,v)$, defined in
Eq.~\eqref{eq:integrated_residual}, as functions of the half-separation
{$d \le 2$ for (a) $v=0.1$, (b) $v=0.2$, and (c) $v=0.5$}. The black solid,
blue dashed, and red dot-dashed curves correspond to the standard
superposition ($A=\mathrm{sup}$), Class~I ($A=\mathrm{I}$), and Class~II
($A=\mathrm{II}$), respectively. {The main semi-logarithmic panels illustrate the global behavior and decay of the residuals across all three
prescriptions. Insets display $\mathcal{R}_{\mathrm{sup}}$ and
$\mathcal{R}_{\mathrm{I}}$ on a linear scale in the strongly nonlinear
regime ($0.1 \le d \le 0.9$), highlighting that the Class~I configuration
significantly reduces the residual (by more than a factor of two at low
velocities) compared to standard superposition.}}
\label{fig:Residual_Integrate_Profiles}
\end{figure*}

For a quantitative global comparison, we evaluate the integrated squared
residual defined in Eq.~\eqref{eq:integrated_residual}.
Figure~\ref{fig:Residual_Integrate_Profiles} shows
$\mathcal{R}_A(d,v)$ as a function of the half-separation {$d \in (0, 2]$ for
$v=0.1$, $0.2$, and $0.5$}. Across this range, the integrated
residuals generally decrease as $d$ increases, reflecting the progressive
reduction of core overlap between the two constituents.

{The main panels display the data on a logarithmic scale to capture the broad
dynamical range and include all three analytical prescriptions simultaneously.
To clearly assess the advantage of the analytical transformation in the strongly
interacting nonlinear regime ($d \lesssim 1$), each panel contains an inset
showing $\mathcal{R}_{\mathrm{sup}}$ and $\mathcal{R}_{\mathrm{I}}$ on a linear scale
for $0.1 \le d \le 0.9$.}

{As evident from the insets, the Class~I configuration offers a pronounced
improvement over standard superposition at small and moderate separations. In
particular, for low velocities ($v=0.1$ and $v=0.2$), $\mathcal{R}_{\mathrm{sup}}$
peaks around $d \approx 0.35$ with values up to two to three times larger than
$\mathcal{R}_{\mathrm{I}}$, demonstrating that the analytical deformation
effectively mitigates the strong localized distortion during the close-approach
phase. For $v=0.5$, the advantage of Class~I emerges and remains prominent in the intermediate separation window from $d \approx 0.3$ up to $d \approx 1.0$,
beyond which both configurations decay exponentially toward zero.}

\section{Conclusion}
\label{sec:conclusion}


We constructed analytical kink-antikink configurations in the $\varphi^4$ 
model by applying a deformation map to exact two-soliton profiles of the 
integrable sine-Gordon theory. The resulting Class~I
and Class~II configurations, together with the conventional
superposition Ansatz, were compared using a common half-separation
parameter $d$. For the deformed configurations, the deformation
parameters were fixed by the matching conditions
\begin{equation}
e^{c_I(d)}
=
\frac{\tanh^2(\gamma d)}{v^2},
\qquad
e^{c_{II}(d)}
=
v^2\tanh^2(\gamma d),
\end{equation}
with $t_0(d)=-d/v$ and
$\gamma=(1-v^2)^{-1/2}$.

None of the mapped configurations is, in general, an exact solution
of the $\varphi^4$ equation of motion. We therefore quantified their
equation-of-motion error using the local residual
$R_A(x;d,v)$ and the integrated squared residual
\begin{equation}
\mathcal{R}_A(d,v)
=
\int_{-\infty}^{+\infty}
\left[R_A(x;d,v)\right]^2\,dx,
\end{equation}
where $A\in\{\mathrm{sup},I,II\}$. The closed-form expressions given
in Appendix~\ref{app:integrated_squared_residual} provide an analytical
basis for evaluating this measure without solving the nonlinear
Klein--Gordon equation numerically.

The displayed profiles and residual plots show that the relative
accuracy of the three Ansatzes depends on both the separation and the
velocity parameter. In the strong-overlap example, the Class~I
configuration has a smaller local residual than the superposition
Ansatz, whereas the Class~II residual exhibits a pronounced central
structure. At larger separations, the local and integrated residuals
of the superposition Ansatz become substantially smaller in the
plotted cases. The integrated-residual curves for $v=0.1$, $0.2$, and $0.5$  also show changes in the relative ordering
of the Ansatzes as $d$ varies.

Thus, within the parameter range
displayed in Fig.~\ref{fig:Residual_Integrate_Profiles}, no single
configuration is uniformly preferred; the observed crossovers are
parameter-dependent features of the analytical residual measures.

The three-dimensional surfaces shown in
Fig.~\ref{fig:Phi3D} are direct evaluations of the prescribed
analytical profiles over the $(x,t)$ domain. They should not be
interpreted as the result of a numerical time evolution of the
$\varphi^4$ field equation. Likewise, the residual analysis does not
constitute a simulation of kink--antikink scattering, collision
outcomes, resonance windows, or oscillon production. The constructed
fields may instead be regarded as mathematically well-defined and
physically motivated candidate profiles for future dynamical studies,
provided that their residuals and subsequent numerical evolution are
examined separately.

A systematic scan of the $(d,v)$ parameter plane, together with direct
numerical time integration of the $\varphi^4$ equation, would be required
to determine the most accurate initial profile for a given collision
protocol and to establish the dynamical consequences of the observed
residual crossovers.

\section*{Acknowledgments}

AMM and DS were supported by the Higher Education and Science Committee of MESCS RA,  grant No. 25IRF/2-1C008. DB was supported by the National Council for Scientific and Technological Development, CNPq, Grants Nos. 402830/2023-7 and 303469/2019-6.


\appendix

\section{Analytical Forms of the Integrated Squared Residual}
\label{app:integrated_squared_residual}

Throughout this Appendix, we use
\begin{equation}
\begin{aligned}
\gamma &= \frac{1}{\sqrt{1-v^2}},
&
q &= \gamma d,
\\
p &= \tanh^2(q),
&
d &> 0,
\qquad 0<v<1 .
\end{aligned}
\label{eq:appendix_parameters}
\end{equation}

For each configuration
$A\in\{\mathrm{Sup},\mathrm{I},\mathrm{II}\}$, the integrated squared
residual is defined as
\begin{equation}
\mathcal{R}_{A}(d,v)
=
\int_{-\infty}^{+\infty}
\left[R_A(x;d,v)\right]^2\,\mathrm{d}x ,
\label{eq:integrated_squared_residual}
\end{equation}
where $R_A(x;d,v)$ is defined by Eq.~\eqref{eq:residual_definition} in the main text.

\subsection{Superposition Configuration}
\label{app:superposition}

For the superposition configuration, the closed-form result is
\begin{equation}
\begin{aligned}
\mathcal{R}_{\mathrm{Sup}}(d,v)
={}&
\frac{24(1-e^{-4q})^2}{\gamma\,\sinh^7(2q)}
\\
&\times
\Big[
6q\cosh(2q)\big(2\cosh^2(2q)+3\big)
\\
&\qquad
-\sinh(2q)\big(11\cosh^2(2q)+4\big)
\Big].
\end{aligned}
\label{eq:closed_form_residual_superposition}
\end{equation}

\subsection{Class~I Configuration}
\label{app:class_I}

The closed-form integrated squared residual for Class~I is
\begin{equation}
\begin{aligned}
\mathcal{R}_{\mathrm{I}}(d,v)
={}&
-\frac{2(p-1)^2}{15(2-p)^{11/2}p^2(1-v^2)^{3/2}}
\\
&\times
\Big[
\sqrt{2-p}\,P_{\mathrm{I}}(p,v)
\\
&\qquad
+15Q_{\mathrm{I}}(p,v)\,\operatorname{arccoth}\!\big(\sqrt{2-p}\big)
\Big].
\end{aligned}
\label{eq:closed_form_residual_class_I}
\end{equation}
Because $0<p<1$, one has $\sqrt{2-p}>1$, so the $\operatorname{arccoth}$ term is real on the physical domain. 

The auxiliary polynomials $P_{\mathrm{I}}(p,v)$ and $Q_{\mathrm{I}}(p,v)$ entering Eq.~\eqref{eq:closed_form_residual_class_I} are given by
\begin{align}
P_{\mathrm{I}}(p,v)
={}&
p^2\big(811 - 1501p + 1414p^2 - 405p^3 + 45p^4\big)
\nonumber\\
&{}+ 6p\big(73 + 32p - 113p^2\big)v^2
\nonumber\\
&{}+ \big(271 - 851p + 524p^2 + 5p^3 + 15p^4\big)v^4,
\label{eq:PI_polynomial}
\end{align}
and
\begin{align}
Q_{\mathrm{I}}(p,v)
={}&
p^2\big(-135 + 330p - 321p^2 
\nonumber\\
&\qquad + 139p^3 - 32p^4 + 3p^5\big)
\nonumber\\
&{}+ 6p(p-1)\big(-3 + 10p + p^2\big)v^2
\nonumber\\
&{}+ (p-1)^2\big(-59 + 14p - 10p^2 + p^3\big)v^4.
\label{eq:QI_polynomial}
\end{align}

\subsection{Class~II Configuration}
\label{app:class_II}

The corresponding closed form for Class~II is
\begin{equation}
\begin{aligned}
\mathcal{R}_{\mathrm{II}}(d,v)
={}&
\frac{2(p-1)^2}{15(1-2p)^{11/2}\big[p(1-v^2)\big]^{3/2}}
\\
&\times
\Bigg[
\sqrt{p(1-2p)}\,P_{\mathrm{II}}(p,v)
\\
&\qquad
-15Q_{\mathrm{II}}(p,v)\,\arctan\!\left(\sqrt{\frac{1-2p}{p}}\,\right)
\Bigg].
\end{aligned}
\label{eq:closed_form_residual_class_II}
\end{equation}
The expression has a branch point at $p=1/2$. For $0<p<1/2$, the principal-branch form is real. For $1/2<p<1$, the physically relevant real-valued expression is obtained by analytic continuation of the principal branch across the cut.

The auxiliary polynomials $P_{\mathrm{II}}(p,v)$ and $Q_{\mathrm{II}}(p,v)$ entering Eq.~\eqref{eq:closed_form_residual_class_II} are given by
\begin{align}
P_{\mathrm{II}}(p,v)
={}&
45-405p+1414p^2-1501p^3+811p^4
\nonumber\\
&{}+ 6p^3\big(-113+32p+73p^2\big)v^2
\nonumber\\
&{}+ p^2\big(15+5p+524p^2-851p^3+271p^4\big)v^4 ,
\label{eq:PII_polynomial}
\end{align}
and
\begin{align}
Q_{\mathrm{II}}(p,v)
={}&
-3+32p-139p^2+321p^3-330p^4+135p^5
\nonumber\\
&{}- 6p^3\big(1+9p-13p^2+3p^3\big)v^2
\nonumber\\
&{}+ p^2(p-1)^2\big(-1+10p-14p^2+59p^3\big)v^4 .
\label{eq:QII_polynomial}
\end{align}


\begin{thebibliography}{99}

\bibitem{manton2004topological}
N.~Manton and P.~Sutcliffe,
{\it Topological Solitons},
\href{https://doi.org/10.1017/CBO9780511617034}{Cambridge University Press, Cambridge U.K. (2010).}

\bibitem{vachaspati2006solitons}
T.~Vachaspati,
{\it Kinks and Domain Walls: An Introduction to Classical and Quantum Solitons},
\href{https://doi.org/10.1017/9781009290456}{Cambridge University Press, Cambridge U.K. (2022).}

\bibitem{Kevrekidis.book.2019}
P.~G.~Kevrekidis,
{\it A Dynamical Perspective on the $\phi^4$ Model, Past, Present and Future},
\href{https://doi.org/10.1007/978-3-030-11839-6}{Springer, Cham (2019)}.

\bibitem{Bishop.PhysD.1980}
A.~R.~Bishop, J.~A.~Krumhansl, and S.~E.~Trullinger,
{\it Solitons in condensed matter: A paradigm},
\href{https://doi.org/10.1016/0167-2789(80)90003-2}{{\it Physica D: Nonlinear Phenomena} {\bf 1} (1980) 1}.

\bibitem{bazeia2003kinks}
D.~Bazeia, J.~Menezes, and R.~Menezes,
{\it  New Global Defect Structures},
\href{https://doi.org/10.1103/PhysRevLett.91.241601}{{\it Phys.\ Rev.\ Lett.} {\bf 91} (2003) 241601}
[\href{https://arxiv.org/abs/hep-th/0305234}{\tt hep-th/0305234}].

\bibitem{Evslin.JHEP.2024}
K.~Ogundipe and J.~Evslin,
{\it Perturbative approach to time-dependent quantum solitons},
\href{https://doi.org/10.1007/JHEP06(2024)174}{{\it JHEP} {\bf 06} (2024) 174}
[\href{https://arxiv.org/abs/2403.13232}{\tt arXiv:2403.13232}].

\bibitem{campbell1983resonance}
D.~K.~Campbell, J.~F.~Schonfeld, and C.~A.~Wingate,
{\it Resonance structure in kink-antikink interactions in $\varphi^4$ theory},
\href{https://doi.org/10.1016/0167-2789(83)90289-0}{{\it Physica D} {\bf 9} (1983) 1}.

\bibitem{Makhankov.PhysRep.1978}
V.~G.~Makhankov,
{\it Dynamics of classical solitons (in non-integrable systems)},
\href{https://doi.org/10.1016/0370-1573(78)90074-1}{{\it Phys.\ Rep.} {\bf 35} (1978) 1}.

\bibitem{rajaraman1982solitons}
R.~Rajaraman,
{\it Solitons and Instantons: An Introduction to Solitons and Instantons in Quantum Field Theory},
North-Holland, Amsterdam (1982).

\bibitem{Dmitriev.Nonlinearity.2000}
A.~E.~Miroshnichenko, S.~V.~Dmitriev, A.~A.~Vasiliev, and T.~Shigenari,
{\it Inelastic three-soliton collisions in a weakly discrete sine-Gordon system},
\href{https://doi.org/10.1088/0951-7715/13/3/318}{{\it Nonlinearity} {\bf 13} (2000) 837}.

\bibitem{Dmitriev.PRE.2008}
S.~V.~Dmitriev, P.~G.~Kevrekidis, and Y.~S.~Kivshar,
{\it Radiationless energy exchange in three-soliton collisions},
\href{https://doi.org/10.1103/PhysRevE.78.046604}{{\it Phys.\ Rev.\ E} {\bf 78} (2008) 046604}.

\bibitem{Moradi.EPJB.2018}
A.~Moradi~Marjaneh, A.~Askari, D.~Saadatmand, and S.~V.~Dmitriev,
{\it Extreme values of elastic strain and energy in sine-Gordon multi-kink collisions},
\href{https://doi.org/10.1140/epjb/e2017-80406-y}{{\it Eur.\ Phys.\ J.\ B} {\bf 91} (2018) 22}
[\href{https://arxiv.org/abs/1710.10159}{\tt arXiv:1710.10159}].

\bibitem{Lohe.PRD.1979}
M.~A.~Lohe,
{\it Soliton structures in $P(\varphi)$ in two dimensions},
\href{https://doi.org/10.1103/PhysRevD.20.3120}{{\it Phys.\ Rev.\ D} {\bf 20} (1979) 3120}.

\bibitem{Campbell.PhysD.1983}
D.~K.~Campbell and M.~Peyrard,
{\it Solitary wave collisions revisited},
\href{https://doi.org/10.1016/0167-2789(86)90161-2}{{\it Physica D} {\bf 18} (1986) 47}.

\bibitem{Gani.EPJC.2018}
V.~A.~Gani, A.~Moradi~Marjaneh, A.~Askari, E.~Belendryasova, and D.~Saadatmand,
{\it Scattering of the double sine-Gordon kinks},
\href{https://doi.org/10.1140/epjc/s10052-018-5813-1}{{\it Eur.\ Phys.\ J.\ C} {\bf 78} (2018) 345}
[\href{https://arxiv.org/abs/1711.01918}{\tt arXiv:1711.01918}].

\bibitem{Saadatmand.EPJB.2022}
D.~Saadatmand and A.~Moradi~Marjaneh,
{\it Scattering of the asymmetric $\varphi^6$ kinks from a $\mathcal{PT}$-symmetric perturbation: creating multiple kink--antikink pairs from phonons},
\href{https://doi.org/10.1140/epjb/s10051-022-00405-x}{{\it Eur.\ Phys.\ J.\ B} {\bf 95} (2022) 144}
[\href{https://arxiv.org/abs/2201.03277}{\tt arXiv:2201.03277}].

\bibitem{Adam.PRD.2022}
C.~Adam, K.~Ole\'s, T.~Roma\'nczukiewicz, and A.~Wereszczy\'nski,
{\it Spectral walls in antikink-kink scattering in the $\phi^6$ model},
\href{https://doi.org/10.1103/PhysRevD.106.105027}{{\it Phys.\ Rev.\ D} {\bf 106} (2022) 105027}
[\href{https://arxiv.org/abs/2209.11479}{\tt arXiv:2209.11479}].

\bibitem{belova1997soliton}
T.~I.~Belova and A.~E.~Kudryavtsev,
{\it Solitons and their interactions in classical field theory},
\href{https://doi.org/10.1070/PU1997v040n04ABEH000227}{{\it Physics-Uspekhi} {\bf 40} (1997) 359}.

\bibitem{Askari.CSF.2020}
A.~Askari, A.~Moradi~Marjaneh, Z.~G.~Rakhmatullina, M.~Ebrahimi-Loushab, D.~Saadatmand, V.~A.~Gani, P.~G.~Kevrekidis, and S.~V.~Dmitriev,
{\it Collision of $\varphi^4$ kinks free of the Peierls-Nabarro barrier in the regime of strong discreteness},
\href{https://doi.org/10.1016/j.chaos.2020.109854}{{\it Chaos, Solitons and Fractals} {\bf 138} (2020) 109854}
[\href{https://arxiv.org/abs/1912.07953}{\tt arXiv:1912.07953}].

\bibitem{sugiyama1979kink}
T.~Sugiyama, {\it Kink-Antikink Collisions in the Two-Dimensional $\phi^4$ Model},
\href{https://doi.org/10.1143/PTP.61.550}{Prog.\ Theor.\ Phys. {\bf 61}, 1550 (1979)}.

\bibitem{anninos1991fractal}
P.~Anninos, S.~Oliveira, and R.~A.~Matzner,
{\it Fractal structure in the scalar $\lambda(\phi^2-1)^2$ theory},
\href{https://doi.org/10.1103/PhysRevD.44.1147}
{Phys.\ Rev.\ D {\bf 44}, 1147 (1991)}.

\bibitem{Manton.PRL.2021}
N.~S.~Manton, K.~Ole\'s, T.~Roma\'nczukiewicz, and A.~Wereszczy\'nski,
{\it Collective Coordinate Model of Kink-Antikink Collisions in $\phi^4$ Theory},
\href{https://doi.org/10.1103/PhysRevLett.127.071601}{{\it Phys.\ Rev.\ Lett.} {\bf 127} (2021) 071601}
[\href{https://arxiv.org/abs/2106.05153}{\tt arXiv:2106.05153}].

\bibitem{Almeida.EPJC.2025}
F.~C.~E.~Lima, R.~Casana, and C.~A.~S.~Almeida,
{\it Kinks and double-kinks in generalized $\phi^4$-and $\phi^8$-models},
\href{https://doi.org/10.1140/epjc/s10052-024-13651-3}{{\it Eur.\ Phys.\ J.\ C} {\bf 84} (2024) 1266}
[\href{https://arxiv.org/abs/2408.04761}{\tt arXiv:2408.04761}].

\bibitem{Saadatmand.CSF.2024}
D.~Saadatmand, A.~Moradi~Marjaneh, A.~Askari, and H.~Weigel,
{\it Phonons scattering off discrete asymmetric solitons in the absence of a Peierls-Nabarro potential},
\href{https://doi.org/10.1016/j.chaos.2024.114550}{{\it Chaos, Solitons and Fractals} {\bf 180} (2024) 114550}
[\href{https://arxiv.org/abs/2308.02322}{\tt arXiv:2308.02322}].

\bibitem{dashen1975semiclassical}
R.~F.~Dashen, B.~Hasslacher, and A.~Neveu,
{\it Particle spectrum in model field theories from semiclassical functional integral techniques},
\href{https://doi.org/10.1103/PhysRevD.11.3424}
{Phys.\ Rev.\ D {\bf 11}, 3424 (1975)}.

\bibitem{Moradi.CNSNS.2017}
A.~Moradi~Marjaneh, D.~Saadatmand, K.~Zhou, S.~V.~Dmitriev, and M.~E.~Zomorrodian,
{\it High energy density in the collision of $N$ kinks in the $\phi^4$ model},
\href{https://doi.org/10.1016/j.cnsns.2017.01.022}{{\it Commun.\ Nonlinear Sci.\ Numer.\ Simulat.} {\bf 49} (2017) 30}
[\href{https://arxiv.org/abs/1605.09767}{\tt arXiv:1605.09767}].

\bibitem{Christov.PRL.2019}
I.~C.~Christov, R.~J.~Decker, A.~Demirkaya, V.~A.~Gani, P.~G.~Kevrekidis, A.~Khare, and A.~Saxena,
{\it Kink-Kink and Kink-Antikink Interactions with Long-Range Tails},
\href{https://doi.org/10.1103/PhysRevLett.122.171601}{{\it Phys.\ Rev.\ Lett.} {\bf 122} (2019) 171601}
[\href{https://arxiv.org/abs/1811.07872}{\tt arXiv:1811.07872}].

\bibitem{Campos.JHEP.2024}
J.~G.~F.~Campos and A.~Mohammadi,
{\it Collisions between kinks with long-range tails: a simple and efficient method},
\href{https://doi.org/10.1007/JHEP02(2024)056}{{\it JHEP} {\bf 02} (2024) 056}
[\href{https://arxiv.org/abs/2309.05628}{\tt arXiv:2309.05628}].

\bibitem{Andre.AnlPhys.2025}
I.~Andrade, M.~A.~Marques, and R.~Menezes,
{\it Analytical short-and long-range kink-like structures in scalar field models with polynomial interactions},
\href{https://doi.org/10.1016/j.aop.2024.169915}{{\it Annals of Physics} {\bf 473} (2025) 169915}
[\href{https://arxiv.org/abs/2409.01961}{\tt arXiv:2409.01961}].

\bibitem{Campos.JHEP.2025}
J.~G.~F.~Campos, A.~Mohammadi, and T.~Roma\'nczukiewicz,
{\it Collective coordinates method for long-range kink collisions},
\href{https://doi.org/10.1007/JHEP01(2025)166}{{\it JHEP} {\bf 01} (2025) 166}
[\href{https://arxiv.org/abs/2411.12630}{\tt arXiv:2411.12630}].

\bibitem{Saadatmand.2014.dynamics}
D.~Saadatmand, A.~Moradi~Marjaneh, and M.~Heidari,
{\it Dynamics of coupled field solitons: A collective coordinate approach},
\href{https://doi.org/10.1007/s12043-014-0797-3}{{\it Pramana -- J.\ Phys.} {\bf 83} (2014) 505}.

\bibitem{Weigel.PRD.2016}
I.~Takyi and H.~Weigel,
{\it Collective coordinates in one-dimensional soliton models revisited},
\href{https://doi.org/10.1103/PhysRevD.94.085008}{{\it Phys.\ Rev.\ D} {\bf 94} (2016) 085008}
[\href{https://arxiv.org/abs/1609.06833}{\tt arXiv:1609.06833}].

\bibitem{Pereira.JoPA.2021}
C.~F.~S.~Pereira, G.~Luchini, T.~Tassis, and C.~P.~Constantinidis,
{\it Some novel considerations about the collective coordinates approximation for the scattering of $\varphi^4$ kinks},
\href{https://doi.org/10.1088/1751-8121/abd815}{{\it J.\ Phys.\ A: Math.\ Theor.} {\bf 54} (2021) 075701}
[\href{https://arxiv.org/abs/2004.00571}{\tt arXiv:2004.00571}].

\bibitem{bazeia2002deformed}
D.~Bazeia, L.~Losano, and J.~M.~C.~Malbouisson,
{\it Deformed defects},
\href{https://doi.org/10.1103/PhysRevD.66.101701}{{\it Phys.\ Rev.\ D} {\bf 66} (2002) 101701(R)}
[\href{https://arxiv.org/abs/hep-th/0209027}{\tt hep-th/0209027}].

\bibitem{bazeia2006deformation}
D.~Bazeia, M.~A.~Gonzalez Leon, L.~Losano, and J.~Mateos Guilarte,
{\it Deformed defects for scalar fields with polynomial interactions},
\href{https://doi.org/10.1103/PhysRevD.73.105008}{{\it Phys.\ Rev.\ D} {\bf 73} (2006) 105008}
[\href{https://arxiv.org/abs/hep-th/0605127}{\tt hep-th/0605127}].

\end{thebibliography}
\end{document}